\documentclass[12pt]{article}
\begin{document}

%\vspace*{1in}

%%%%%%%%%%%%%%%%%%%%%%%%%%%%%%%%%%%%%%%%%%%%%%%%%%%%%%%%%%%%%%%%%%%%%%%%%%%%%%%%
\title {A new approach to the analysis of a noncommutative Chern--Simons theory}
\author
{Pradip Mukherjee \footnote{pradip@bose.res.in} {}\footnote{Also Visiting Associate, S. N. Bose National Centre 
for Basic Sciences, JD Block, Sector III, Salt Lake City, Calcutta -700 098, India}
and Anirban Saha\\
Department of Physics, Presidency College\\
86/1 College Street, Kolkata - 700 073, India}
\maketitle
\abstract
{A novel approach to the analysis of a noncommutative Chern--Simons gauge
theory with matter coupled in the adjoint representation has been discussed.
The analysis is based on a recently proposed closed form Seiberg--Witten map
which is exact in the noncommutative parameter. }

\noindent {\bf PAC codes:} {11.10.Nx, 11.15-q}
   
The idea of fuzzy space time where the coordinates $x^{\mu}$ satisfy the
noncommutative (NC) algebra 
\begin{equation}
\left[x^{\mu}, x^{\nu}\right] = i \theta^{\mu \nu}
\label{ncgometry}
\end{equation}
where $\theta^{\mu \nu}$ is a constant anti-symmetric tensor, was mooted
long ago \cite{sny}. This idea has been revived in the recent past and
field theories defined over this NC space %\cite{szaboref} 
are currently the subject of very intense research \cite{szabo}.  One
approach of analysis of the NC field theories is to work in a
certain Hilbert space which carries a representation of the basic
NC algebra. The fields are defined in this Hilbert space by the
Weyl--Wigner correspondence. The operator approach
is easily extended to the abelian and nonabelian gauge groups
\cite{mad}.  An alternative approach of treating NC theories is to work in 
the deformed phase space where the ordinary product is replaced by the star product. 
In this formalism the fields are defined as functions of the phase space variables with 
the product of two fields $\hat \phi(x)$ and $\hat \psi(x)$ given by the star product
\begin{equation}
\hat \phi(x) \star \hat \psi(x) = \left(\hat \phi \star \hat \psi \right)(x) = e^{\frac{i}{2}
\theta^{\alpha\beta}\partial_{\alpha}\partial^{'}_{\beta}} 
  \hat \phi (x) \hat \psi(x^{'})\big{|}_{x^{'}=x.} 
\label{star}
\end{equation}

         An important breakthrough in the investigation of the NC gauge
theories has
been achieved by Seiberg and Witten \cite{SW} from their study of the
Dirac--Born--Infeld (DBI) action of open string dynamics on a D-Brane
obtained in the limit of slowly varying fields \cite{frad}.
It was observed that depending on
the regularization scheme one can have alternative descriptions
of the theory in terms of commutative and noncommutative models. Since physics
must be independent of the particular regularization scheme a space time
redefinition between the ordinary and noncommutative gauge fields is
indicated.

    Prompted by Seiberg and Witten's seminal work a new approach towards the study of NC  gauge theories has originated in the literature both for the abelian \cite{bichl} and non-abelian \cite{ban}  gauge groups where the NC gauge theories have been analysed from their commutative equivalent counterparts. Various NC models have been analysed in the recent past from this point of view \cite{all}.  The essence of this approach is to formulate the theory on the phase space and expand the star products (\ref{star}) with appropriate Seiberg--Witten maps implemented individually on the fields of the model. Consequently, the outcome is in the form of a perturbative expansion in the noncommutative parameter. It will be very much desirable if this analysis can be done in a closed form such
that results exact to all orders in $\theta$ are obtained. Naturally, the
possibility of this rests on the availabilibility of the SW maps in a closed
form.

   Recently, a method of obtaining SW maps for certain models has been devised which is exact in 
the NC parameter \cite{hyun,rbhsk}. This is based on the change of variables between open and closed 
string parameters and connection of the approach with the deformation quantization technique \cite{Kont} 
has been
demonstrated \cite{rbhsk}. Specifically, an exact map for an adjoint scalar field has been found \cite{rbhsk}, consistent with that deduced from RR couplings of unstable non-BPS D-branes \cite{mukhi}.  In the present letter we will use this map to analyze a $U(1)_{\star}$ Chern--Simons (C--S) coupled scalar
field theory in $2+1$ dimensional flat space time where the scalar field is in the adjoint
representation of the gauge group. Models with the NC scalar field in the adjoint representation have been considered earlier from the operator approach with the gauge field dynamics governed solely by the Maxwell term \cite{poly} and also by a combination of the Maxwell and the C--S term \cite{ prnjp}. Our selection of the model is motivated by the fact that in the commutative limit the scalar field decouples from the gauge interaction. In other words any non-trivial result of our analysis comes from the NC features only. For the same reason we chose the C--S coupling because it remains form-invariant under SW map \cite{gs}.
Apart from this the C--S theories have been studied both in the commutative \cite{our} and noncommutative settings \cite{suskind} principally due to their inherent interest in connection with the theory of fractional spin and statistics.

 The action of our theory is given by
\begin{eqnarray}
\hat S &=& \int d^{3}x\left[\frac{1}{2} \left(\hat D_{\mu} \star \hat \phi \right)\star \left(\hat D^{\mu} \star \hat \phi\right) + \frac{k}{2}\epsilon^{\mu \nu \lambda}\left(\hat A_{\mu} \star \partial_{\nu}\hat A_{\lambda} - \frac{2i}{3} \hat A_{\mu} \star \hat A_{\nu} \star \hat A_{\lambda}\right)\right]
\label{ncaction}
\end{eqnarray}
where $\hat \phi$ is the scalar field and $\hat A_{\mu}$ is the NC C--S gauge
field. We adopt the Minkowski metric $\eta_{\mu \nu} = {\rm diag} \left( +,-,-,-\right)$. The covariant derivative $\hat D_{\mu}\star \hat \phi$ is defined as
\begin{equation}
\hat D_{\mu}\star \hat \phi = \partial_{\mu} \hat \phi - i \left[\hat A_{\mu}, \phi\right]_{\star}
\label{covder}
\end{equation}
The action (\ref{ncaction}) is invariant under the $\star$-gauge
transformation
\begin{equation}
\hat \delta_{\hat \lambda} \hat A_{\mu} = \hat D_{\mu} \star \hat \lambda, \qquad \hat \delta_{\hat \lambda} \hat \phi = -i \left[\hat \phi, \hat \lambda  \right]_{\star}
\label{starg}
\end{equation}

   The commutative version of (\ref{ncaction}) can easily be obtained by using the exact SW map for $\hat D_{\mu}\star\hat \phi(x)$ given in \cite{rbhsk} and noting that the C--S action retains its form under SW map. Proceeding in this direction we write the commutative equivalent of (\ref{ncaction}) as
\begin{eqnarray}
\hat S &\stackrel{\rm{SW \; map}}{=}&  \int d^{3}x \left[\frac{1}{2}\sqrt{{\mathrm {det}} \left( 1 + F \theta \right)}\left(\frac{1}{1 + F \theta }\frac{1}{1 + \theta F}\right)^{\mu \nu} 
\partial_{\mu} \phi\partial_{\nu} \phi + \frac{k}{2}\epsilon^{\mu \nu \lambda} A_{\mu} \partial_{\nu} A_{\lambda}\right] 
\label{caction}
\end{eqnarray}
%The first term in (\ref{caction}) can be written as an ordinary scalar coupled to a field dependent metric \cite{rbhsk} and the C--S term, being metric inependent, does not change its form.
In (\ref{caction}) we have used the matrix notation
\begin{eqnarray}
\left(AB\right)^{\mu \nu} = A^{\mu \lambda}B_{\lambda}{}^{\nu} 
%AB = A^{\mu \lambda}B_{\lambda}{}_{\mu}
\label{matnot}
\end{eqnarray}
Also $\left( 1 + F \theta \right)$ is to be interpreted as a mixed tensor in calculating the determinant. Note that the quartic term in the C--S action vanishes in the commutative equivalent version. The scalar field part of the action (\ref{caction}) can be written as an ordinary scalar field theory coupled with a gravitational field induced by the dynamical gauge field. However, the dynamics of the gauge field, being dictated by the Chern--Simons three-form, is unaffected by the induced gravity. If we would instead consider Maxwell theory then the coupling should equally affect the gauge field dynamics also \cite{rbhsk}.

   From (\ref{caction}) we readily observe that in the commutative limit ($\theta_{\mu \nu} \to 0$) the gauge field decouples, leading to the well known fact that there is no non-trivial gauge coupling
of the neutral scalar field in the corresponding commutative field theory. The commutative equivalent to the transformations (\ref{starg}) are 
\begin{equation}
\delta_{\lambda} A_{\mu} =\partial_{\mu}\lambda, \qquad \delta_{\lambda} \phi = 0
\label{g}
\end{equation}
Clearly, the action (\ref{caction}) is manifestly invariant under (\ref{g}). 

       It is now straightforward to write down the equations of motion
for  the scalar field $\phi$ and the gauge field $A_{\mu}$ from
(\ref{caction}) respectively as 
\begin{equation}
\partial_{\alpha}\left\{\sqrt{{\mathrm {det}} \left( 1 + F \theta \right)}\left(\frac{1}{1 + F \theta }\frac{1}{1 + \theta F}\right)^{\alpha \nu} \partial_{\nu} \phi\right\} = 0
\label{eqmphi}
\end{equation}
and
\begin{equation}
k\epsilon^{\alpha \nu \lambda} \partial_{\nu} A_{\lambda} = j^{\alpha}
\label{eqmA}
\end{equation}
where, in (\ref{eqmA}), 
\begin{eqnarray}
j^{\alpha} &=& \partial_{\xi}\left[\sqrt{{\mathrm {det}} \left( 1 + F \theta \right)} \left\{\frac{1}{4} \left(\theta\frac{1}{1 + F \theta } + \frac{1}{1 + \theta F} \theta\right)^{\alpha \xi} \left(\frac{1}{1 + F \theta }\frac{1}{1 + \theta F}\right)^{\mu \nu} \right.\right.\nonumber\\ 
&&+
\left(\frac{1}{1 + F \theta }\frac{1}{1 + \theta F} \theta \right)^{\mu \alpha} \left(\frac{1}{1 + \theta F}\right)^{\xi \nu} + \left. \left. \left(\frac{1}{1 + F \theta}\right)^{\mu \alpha} \left(\theta \frac{1}{1 + F \theta }\frac{1}{1 + \theta F} \right)^{\xi \nu}
\right\}
\partial_{\mu} \phi\partial_{\nu} \phi\right]\nonumber \\
\label{defJ}
\end{eqnarray}
Certain observations about the above equations are in order.
In the commutative limit or (and) vanishing gauge field
\begin{equation}
\left(\frac{1}{1 + F \theta }\frac{1}{1 + \theta F}\right)^{\mu \nu} \to \eta^{\mu \nu} 
\label{mat}
\end{equation}
Thus the equation of motion for $\phi$ in (\ref{eqmphi}) reduces to the expected
form $\partial_{\mu}\partial^{\mu}\phi = 0$ in these limits.
Again going to the commutative limit we find that the gauge field equation
becomes trivial, which is also a characteristic feature of C--S theories
without any matter coupling.
 %\cite{dunne} 
By direct computation, we get from (\ref{defJ}) 
\begin{equation}
\partial_{\alpha} j^{\alpha}= 0
\label{cont}
\end{equation}
This exhibits the consistency of (\ref{eqmA}).
Naturally, $j^{\alpha}$ is interpreted as the matter current. 

     At this point it can be noted that the usual approach of obtaining the
commutative equivalent of (\ref{ncaction}) is to expand the star products and use separate maps for the gauge fields and matter fields in the form of perturbative expansions in the NC parameter $\theta$ \cite{all}. To the lowest order in $\theta$ the explicit forms of the SW maps are known as \cite{SW,bichl}
\begin{eqnarray}
\hat \psi &=& \psi - \theta^{mj}A_{m}\partial_{j}\psi \nonumber\\
\hat A_{i} &=& A_{i} - \frac{1}{2}\theta^{mj}A_{m}
\left(\partial_{j}A_{i} + F_{ji}\right)
\label{1stordmp}
\end{eqnarray}
Using these expressions and the star product (\ref{star}) to order $\theta$ in (\ref{ncaction}) we get
\begin{eqnarray}
\hat S &\stackrel{\rm{SW \; map}}{=}&  \int d^{3}x \left[\left\{\frac{1}{2} \partial^{\mu} \phi \partial_{\mu} \phi - \theta ^{\alpha \beta} F^{\mu}{}_{\alpha} \partial_{\beta} \phi \partial_{\mu} \phi - \theta ^{\alpha \beta} A_{\alpha}\partial_{\mu} \partial_{\beta} \phi \partial^{\mu} \phi\right\} + \frac{k}{2}\epsilon^{\mu \nu \lambda} A_{\mu} \partial_{\nu} A_{\lambda}\right] \nonumber\\
\label{1stordac}
\end{eqnarray}
We can show explicitly that the first order approximation of (\ref{caction}) matches exactly with (\ref{1stordac}). Naturally, the equations of motion (\ref{eqmphi}) and (\ref{eqmA}) should agree upto the first order with those following from the conventional first order action (\ref{1stordac}). Expanding (\ref{eqmphi}) and (\ref{eqmA}) to first order in $\theta$ we get
\begin{eqnarray}
\partial_{\alpha}\left[\left\{1 + {\mathrm Tr} \left( F \theta \right)\right\} \partial^{\alpha} \phi - \left(F\theta + \theta F\right)^{\alpha \nu} \partial_{\nu} \phi\right] = 0 
\label{eqmphi1}
\end{eqnarray}
and
\begin{eqnarray}
k\epsilon^{\alpha \nu \lambda} \partial_{\nu} A_{\lambda} &=& \partial_{\xi} [ \frac{1}{2} \theta^{\alpha \xi} \partial_{\mu} \phi \partial^{\mu} \phi + \theta^{\mu \alpha} \partial_{\mu} \phi \partial^{\xi} \phi + \theta^{\xi \mu} \partial^{\alpha} \phi \partial_{\mu} \phi ] 
\label{eqma1}
\end{eqnarray}
respectively. One can verify easily that the same equations follow as Euler--Lagrange equations from (\ref{1stordac}).

   We now turn to the construction of an energy momentum (EM) tensor of our model (\ref{caction}). The issue of energy momentum tensor for a noncommutative gauge theory involves many subtle points as evidenced in the literature \cite{rbyl}. It is thus instructive to address the question from different approaches, which in the context of commutative models are known to lead to equivalent conclusions but the same is not true {\it apriori} for NC gauge theories. Indeed, the commutative equivalent model offers an appropriate platform to discuss these aspects. 

    We begin with the construction of the Noether EM tensor. Consider the infinitesimal space time translation $x^{\mu}  \to \left(x^{\mu}+ a^{\mu}\right)$ under which the fields $\phi$ and $A_{\mu}$ transform as 
\begin{eqnarray}
\delta \phi = a^{\mu}\partial_{\mu}\phi, \qquad \delta A_{\mu} = a^{\nu}\partial_{\nu} A_{\mu}
\label{trans}
\end{eqnarray}
From the invariance of the theory we get the following form of the EM tensor in the usual way, 
%---------------------------
%THE CANONICAL EMT(exact):
%------------------------
\begin{eqnarray}
\Theta^{c}_{\rho \sigma} &=& \sqrt{{\mathrm det} \left( 1 + F \theta \right)}\left[\left(\frac{1}{1 + F \theta }\frac{1}{1 + \theta F}\right)_{\rho}{}^{\nu}\partial_{\nu} \phi \partial_{\sigma} \phi\right.\nonumber\\
&&+ \left\{\frac{1}{4} \left(\theta\frac{1}{1 +F \theta} + \frac{1}{1 + \theta F} \theta\right)^{\alpha}{}_{\rho}\left(\frac{1}{1 + F \theta }\frac{1}{1 + \theta F}\right)^{\mu \nu}\right.\nonumber\\
&&+
\left(\frac{1}{1 + F \theta }\frac{1}{1 + \theta F} \theta \right)^{\mu \alpha} \left(\frac{1}{1 + \theta F}\right)_{\rho}{}^{\nu} \nonumber\\
&&+  \left. \left(\frac{1}{1 + F \theta}\right)^{\mu \alpha} \left(\theta \frac{1}{1 + F \theta }\frac{1}{1 + \theta F} \right)_{\rho}{}^{\nu}
\right\} \left( \partial_{\mu} \phi\partial_{\nu} \phi\right)\left( \partial_{\sigma} A_{\alpha} \right) \nonumber\\
&&- \left. \frac{1}{2} \eta_{\rho \sigma} \left(\frac{1}{1 + F \theta }\frac{1}{1 + \theta F}\right)^{\mu \nu} \partial_{\mu} \phi\partial_{\nu} \phi\right]\nonumber\\
&&- \frac{k}{2}\left(\epsilon_{\rho}{}^{\mu \alpha} A_{\mu} \partial_{\sigma} A_{\alpha} + \eta_{\rho \sigma} \epsilon^{\mu \nu \alpha} A_{\mu} \partial_{\nu} A_{\alpha}\right)
\label{noetheremt}
\end{eqnarray}
%----------------------------------------------------------------------------------------------------------------------------
%THE FIRST ORDER CANONICAL EMT:
%------------------------------
%\begin{eqnarray}\Theta^{c}_{\rho \sigma} =\{1+\frac{1}{2}\rm{Tr}\left(F\theta\right)\}\left[\partial_{\rho}\phi\partial_{\sigma}\phi - \frac{1}{2}\eta_{\rho\sigma}\partial_{\mu}\phi\partial^{\mu}\phi\right] - \frac{k}{2}\left(\epsilon_{\rho}{}^{\mu \alpha} A_{\mu} \partial_{\sigma} A_{\alpha} + \eta_{\rho \sigma} \epsilon^{\mu \nu \alpha} A_{\mu} \partial_{\nu} A_{\alpha}\right)
%\end{eqnarray}
%----------------------------------------------------------------------------------------------------------------------------
The Noether E--M tensor is useful to construct the generators of space time transformations. However, it is neither gauge invariant nor symmetric. One would then like to improve it to get a gauge invariant EM tensor using Belinfante's method. A better alternative is to considr a subsequent gauge transformation with the spatial translation (\ref{trans}) so that the gauge field transform covariantly, 
\begin{eqnarray}
\delta A_{\mu} = a^{\nu}F_{\nu\mu}
\label{trans1}
\end{eqnarray}
and obtain an improved EM tensor by Noether's method \cite{jackiw} using the modified transformation. This leads to 
%------------------------
%THE JACKIW-PI EMT(exact):
%------------------------
\begin{eqnarray}
{\mathcal T}_{\rho \sigma} &=& \frac{\partial {\mathcal L}}{\partial\left(\partial^{\rho}\phi\right)}\partial_{\sigma}\phi +  \frac{\partial {\mathcal L}}{\partial\left(\partial^{\rho}A_{\alpha}\right)}F_{\sigma\alpha} - \eta_{\rho\sigma}{\mathcal L}\nonumber\\
&&=\sqrt{{\mathrm det} \left( 1 + F \theta \right)}\left[\left(\frac{1}{1 + F \theta }\frac{1}{1 + \theta F}\right)_{\rho}{}^{\nu}\partial_{\nu} \phi \partial_{\sigma} \phi\right.\nonumber\\
&&+ \left\{\frac{1}{4} \left(\theta\frac{1}{1 +F \theta} + \frac{1}{1 + \theta F} \theta\right)^{\alpha}{}_{\rho}\left(\frac{1}{1 + F \theta }\frac{1}{1 + \theta F}\right)^{\mu \nu}\right.\nonumber\\
&&+
\left(\frac{1}{1 + F \theta }\frac{1}{1 + \theta F} \theta \right)^{\mu \alpha} \left(\frac{1}{1 + \theta F}\right)_{\rho}{}^{\nu} \nonumber\\
&&+  \left. \left(\frac{1}{1 + F \theta}\right)^{\mu \alpha} \left(\theta \frac{1}{1 + F \theta }\frac{1}{1 + \theta F} \right)_{\rho}{}^{\nu}
\right\} \left( \partial_{\mu} \phi\partial_{\nu} \phi\right)\left(F_{\sigma\alpha} \right) \nonumber\\
&&- \left. \frac{1}{2} \eta_{\rho \sigma} \left(\frac{1}{1 + F \theta }\frac{1}{1 + \theta F}\right)^{\mu \nu} \partial_{\mu} \phi\partial_{\nu} \phi\right]\nonumber\\
&&- \frac{k}{2}\left(\epsilon_{\rho}{}^{\mu \alpha} A_{\mu} F_{\sigma\alpha} + \eta_{\rho \sigma} \epsilon^{\mu \nu \alpha} A_{\mu} \partial_{\nu} A_{\alpha}\right)
\label{jackiwEMT}
\end{eqnarray}
Apart from the contribution from the C--S part this expression is gauge invariant. However, it is not symmetric. In the commutative theories this part of the improved EM tensor becomes simultaniously symmetric. 
%and traceless. 
The exception in the context of NC gauge theories has already been mentioned and is due to the fact that Lorentz and classical conformal invariance are broken in such theories \cite{rbyl}. 
We have observed that the canonical procedures do  not lead to a satisfactory EM tensor. An alternative procedure is to vary the action (\ref{caction}) with respect to a background metric and finally keeping the metric flat. We thus extend the action (\ref{caction}) as 
\begin{eqnarray}
S = \int d^{3}x \sqrt{- g} \mathcal{L}
\label{lag}
\end{eqnarray}
where $g = \mathrm{det} g_{\mu \nu}$ and  $g_{\mu \nu}$ is the background metric. The pure C--S part of (\ref{caction}) is generally covariant irrespective of any metric. Thus the Lagrangean $\mathcal{L}$ in (\ref{lag}) is taken to be the Lagrangean of (\ref{caction}) without the C--S kinetic term. 
%Also note that under passive Lorentz transformation $\theta^{\mu \nu}$ transforms as a second rank tensor and there is no problem in writing $S$ in the covariant form (\ref{lag}) with respect to an arbitrary background metric. 
The EM tensor is obtained from 
\begin{eqnarray}
\Theta^{\left(s\right)}_{\alpha\beta} = 2 \frac{\partial \mathcal{L}}{\partial g^{\alpha \beta}} - \mathcal{L} g_{\alpha \beta}
\label{SEMT}
\end{eqnarray}
in the limit $g_{\mu \nu} \to \eta_{\mu \nu}$. Explicitly
\begin{eqnarray}
\Theta^{\left(s\right)}_{\alpha\beta} &=& \frac{1}{2} \sqrt{{\mathrm {det}} \left( 1 + F \theta \right)} \left[\frac{1}{2}\left(\theta F \frac{1}{1 + \theta F} + \frac{1}{1 + F \theta} F \theta\right)_{\alpha \beta} \left(\frac{1}{1 + F \theta }\frac{1}{1 + \theta F}\right)^{\mu \nu} \partial_{\mu} \phi \partial_{\nu} \phi\right. \nonumber\\
&&+ \left.\left(\frac{1}{1 + F \theta }\frac{1}{1 + \theta F}\right)_{\alpha}{}^{\nu} \partial_{\beta} \phi \partial_{\nu} \phi+ \left(\frac{1}{1 + F \theta }\frac{1}{1 + \theta F}\right)_{\beta}{}^{\nu}\partial_{\alpha} \phi \partial_{\nu} \phi\right] -  \mathcal{L} \eta_{\alpha \beta}\nonumber\\
\label{symemt}
\end{eqnarray}
Note that by construction this EM tensor is both symmetric and gauge invariant. We can conclude that of the various expressions given above this form is the most satisfactory and can be identified as the physical EM tensor. 

                The equations (\ref{eqmphi}) and (\ref{eqmA}) are a set of coupled nonlinear equations. It will thus be instructive to investigate whether they admit any solitary wave solution. This can be seen in a systemetic way by looking for the Bogomolnyi bounds of the equations. To this end we require the energy functional which can appropriately be constructed from the physical EM tensor (\ref{symemt}).   
                  Note that until now our approach was completely general. Specifically, we did not assume vanishing time-space noncommutativity i.e. $\theta^{0i} = 0$. The issue of non zero time space noncommutativity is an involved subject in the literature. It has been argued that noncommutativity in this sector spoils unitarity \cite{gomis} and causality \cite{sei} but there also exists counter examples \cite{bala}. However, assuming $\theta^{0i} = 0$ is almost conventional in the study of NC solitons and in the context of odd dimensional theories it is always possible to do so.  A la this tradition we now assume that the noncommutativity exists only in the spatial direction. Going over to this limit the energy functional becomes
\begin{eqnarray}
E &=& \int d^{2}x \Theta^{\left(s\right)}_{00} = \int d^{2}x \frac{1}{2} \sqrt{{\mathrm{det}} \left( 1 + F \theta \right)}\left\{2\left(\frac{1}{1 + F \theta }\frac{1}{1 + \theta F}\right)_{0}{}^{\nu} \left( \partial_{0} \phi \partial_{\nu} \phi\right)\right.\nonumber\\ && - \left.\left(\frac{1}{1 + F \theta }\frac{1}{1 + \theta F}\right)^{\mu \nu} \left( \partial_{\mu} \phi \partial_{\nu} \phi\right)\right\}\nonumber\\
\label{statenrgy}
\end{eqnarray}
With the stated assumptions about NC tensor $\theta^{\mu \nu}$ the form of the matrices appearing in the above equation can be easily worked out. Explicitly, the matrix for $\left(\frac{1}{1 + F \theta }\frac{1}{1 + \theta F}\right)^{\mu \nu}$ can be written as
\begin{eqnarray}
\left(\begin{array}{ccc}
\left\{1 - \frac{\theta^{2}\left(E_{1}^{2} + E_{2}^{2}\right)}{\left(1 - \theta B\right)^{2}}\right\} & \frac{\theta E_{2}}{\left(1 - \theta B\right)^{2}}  & \frac{-\theta E_{1}}{\left(1 - \theta B\right)^{2}}  \\
\frac{\theta E_{2}}{\left(1 - \theta B\right)^{2}}  & \frac{-1}{\left(1 - \theta B\right)^{2}}  & 0 \\
\frac{-\theta E_{1}}{\left(1 - \theta B\right)^{2}} & 0 & \frac{-1}{\left(1 - \theta B\right)^{2}}\end{array}\right)
\label{thematrix}
\end{eqnarray}
Also ${\rm det}\left(1 + F\theta \right) = \left(1 - \theta B\right)^{2}$. While extracting the square root of this determinant one has to take positive value only. So for $\theta B < 1$, $\sqrt{{\rm det}\left(1 + F\theta \right)} = \left(1 - \theta B\right)$ whereas for $\theta B > 1 $ it is to be replaced by $\left(\theta B - 1 \right)$. The critical point $\theta B = 1$ is known to be a general feature of the NC models, the origin of which can be traced back to noncommutativity in planar quantum mechanics \cite{ban1}. 

 The static limit of the energy functional will now be worked out. First, we observe from (\ref{defJ}) that for $\theta_{0i} = 0$, $j^{0}$ vanishes in the static limit. This leads to vanishing $B$-field, as can be seen from (\ref{eqmA}), making the coupling trivial. The expression of the energy functional (\ref{statenrgy}) becomes, 
\begin{equation}
E = \int d^{2}x  \left\{ \left( \partial_{1} \phi \right)^{2} + \left( \partial_{2} \phi \right)^{2} \right\}\label{statenrgyexpli}
\end{equation}
The energy functional is  positive definite and trivially minimized. Clearly, there is no non-trivial solutions. We thus observe that there is no BPS soliton of the model. Note that nontrivial soliton solutions has been found in NC adjoint scalar field theories \cite{poly,prnjp} with Maxwell coupling. However, these soliton solutions become singular in the $\theta \mapsto 0$ limit. Since our approach has a smooth commutative limit, based as it is on the SW map, such singular solutions (if any) are not included in our model.
    
We have discussed a novel approach of analysing a Chern--Simons (C--S)
coupled real scalar field theory. A commutative equivalent of the model is
obtained which is exact in the noncommutative (NC) parameter $\theta$. This is based on a recently proposed exact Seiberg--Witten (SW) map \cite{rbhsk} which does not require explicit expansion of $\star$-product. At this point the approach is markedly different from the  usual analysis of NC gauge theories using SW fields where one uses the maps in the form of series expansions in $\theta$ \cite{bichl, ban} along with an expansion of the $\star$-product of the functions. Equations of motion satisfied by the dynamical fields have been written down without any restriction on the noncommutative structure. The resulting matter current has been shown to be conserved by explicit calculation which again provides a
consistency check of our equations of motion. We have also demonstrated that
upto first order in the NC parameters our commutative equivalent action
is mapped into the usual version. Different forms of the energy momentum (EM) tensor have been worked out. It was observed that a satisfectory EM tensor can not be obtained from the canonical prescriptions. A symmetric {\it and} gauge invariant EM tensor is constructed by varying the action with respect to the a background metric and this has been identified as the physical EM tensor for our model. Specializing this NC tensor by neglecting noncommutativity in the time--space direction we have shown that the model does not have any nontrivial BPS soliton. Note that this only indicates the absence of such solutions in the sector which has a smooth commutative limit and any other singular soliton solution is not ruled out. The NC Maxwell term can be straightforwardly added in our model which will also be useful in the context of comparision with known results. Also the exact commutative equivalent approach illustrated here may be extended to scalar fields in the fundamental representation if the corresponding exact SW map can be devised. This and other related issues will be taken up subsequently. 

\section* {Acknowledgment}
{PM likes to thank the Director, S. N. Bose National Centre for Basic Sciences for the award of visiting associateship. AS wants to thank the Council of Scientific and Industrial Research (CSIR), Govt. of India, for financial support and the Director, S. N. Bose National Centre for Basic Sciences, for providing computer facilities.}

\end{document}